# Quantum key distribution based on phase encoding and polarization measurement

Hai-Qiang Ma, Jian-Ling Zhao, and Ling-An Wu

Laboratory of Optical Physics, Institute of Physics,

Chinese Academy of Sciences, Beijing 100080, China

A one-way quantum key distribution scheme based on intrinsically stable Faraday-mirror type Michelson interferometers with four-port polarizing beampslitters has been demonstrated which can compensate for birefringence effects automatically. The encoding is performed with phase modulators, but decoding is accomplished through measurement of the polarization state of Bob's photons. An extinction ratio of about 30dB was maintained for several hours over 50km of fiber at 1310nm without any adjustment to the setup, which shows its good potential for practical systems.

As a means for establishing a cryptographic key between two parties Alice and Bob, quantum key distribution (QKD) has gained increasing importance because its security is guaranteed by quantum mechanics, such that any measurement or cloning of quantum information is impossible without disturbing the system. Since the first demonstration by Bennett and co-workers,[1] many schemes have been proposed to realize stable practical QKD systems over fiber[2-5] or free space.[6,7] These differ according to the protocol used，the type of phase or polarization encoding, one-way or round-trip propagation, single-photon or continuous quantum field information carrier, and so forth. The fiber systems most widely adopted now are based on phase encoded single photons that are transmitted through two unbalanced Mach-Zehnder interferometers.[2] However, a one-way system has problems of alignment and compensation for birefringence because the pulses of the two interfering beams do not follow the same path within the two interferometers. To overcome these shortcomings, Gisin et al[4] devised a round-trip plug-and-play system based on Faraday mirrors in which the pulses travel the same path so that fiber birefringence is compensated for automatically, and high-quality interference is obtained. The disadvantage, however, is the low efficiency because each signal has to travel back and forth along the transmission line, which doubles fiber losses. Furthermore, the scheme is vulnerable to the Trojan horse attack from an eavesdropper.[8] A very stable one-way scheme with automatic birefringence compensation was recently demonstrated over more than 125km of trunk fiber, but its efficiency was reduced because photons are wasted at the four-way couplers out of the interferometers.[9]

In this letter we describe an extremely stable one-way fiber QKD scheme with automatic birefringence compensation that has a fourfold increase in efficiency compared with previous methods. It is based on phase encoding and polarization measurement through use of a phase modulator and four-port polarization beamsplitter/coupler in a Faraday mirror (FM) type Michelson interferometer for both Alice and Bob. Compared to other fiber systems, it is simple, easy to operate, and can be used with both BB84 and B92 protocols.[2]

The setup for communication between Alice and Bob is sketched in Fig. 1. Light from a pulsed laser in the +45° polarization state is attenuated to single-photon level by a variable optical attenuator A to prevent possible attacks from Eve,[10] and then equally divided into two orthogonally polarized pulses, $\Lambda_1$ and $\Lambda_2$, by the polarization beamsplitter PBS1. For convenience we define that PBS1 reflects (transmits) the vertically (horizontally) polarized pulse $\Lambda_1$ ($\Lambda_2$) in the states $|v\rangle = \begin{bmatrix} 0 \\ 1 \end{bmatrix}$, $|h\rangle = \begin{bmatrix} 1 \\ 0 \end{bmatrix}$. The former enters the short arm $S_A$ of a Michelson interferometer where it is modulated by a polarization insensitive phase modulator PM1 then reflected back and converted to the $|h\rangle$ state by a 90° rotation Faraday mirror FM1 so that on return it is transmitted through PBS1. The horizontally polarized beam $|h\rangle$ originally transmitted by PBS1 propagates along the long arm ($L_A$) through a delay line DL1 and is reflected back by the Faraday mirror FM2 to return in the $|v\rangle$ state so that it is reflected out by PBS1. The two orthogonally polarized beams thus travel together through the quantum channel QC to an identical Michelson interferometer in Bob's security region. The polarization controller PC is used to correct for birefringence effects introduced by the quantum channel.[5, 11] On arriving at the polarization beamsplitter PBS2 an $|h\rangle$ state photon is transmitted and coupled into the long arm of Bob's interferometer through a delay line DL2 to be reflected back by FM4 in the $|v\rangle$ state, so that on return at PBS2 it is reflected out of the interferometer. The $|v\rangle$ state photon from Alice is reflected by PBS2 into the short arm $S_B$, modulated by the phase modulator PM2, reflected by FM3 to return to PBS2 in the $|h\rangle$ state, and is finally transmitted through PBS2, thus exiting collinearly with the other output beam. When the optical path lengths traversed by the two orthogonally polarized pulses $\Lambda_1$ and $\Lambda_2$ in the two Michelson interferometers, denoted by $S_A+L_B$ and $L_A+S_B$, are exactly equal they re-enter PBS2 at the same time and so combine into one pulse with a polarization determined by the relative phase shifts introduced by Alice and Bob in their phase modulators. Finally, the state of the photon is measured by a polarization projection scheme which depends on whether the BB84 or B92 protocol[2] is chosen. Bob communicates to Alice over a public channel the measurement basis that he used to measure each photon, then establish their secret key. The essential difference of this QKD scheme is that the bit information is encoded on the phase of the photons, but is extracted through measurement of the polarization.

We now present an analysis of the polarization states during the QKD process. The transfer function of a round trip along a birefringent fiber element terminated by a Faraday rotator mirror can be expressed as:[9,12]

$$T = \tilde{T}_E \bullet T_{FM} \bullet \vec{T}_E = \exp(i\beta) T_{FM} \qquad (1)$$

where $\vec{T}_E$ and $\tilde{T}_E$ are the forward and backward Jones transform matrices along the fiber, $T_{FM}$ is the FM's transform matrix, and $\beta$ is a phase shift due to birefringence effects in the fiber. Thus for a FM containing a 45° rotator the return polarization is always purely orthogonal to the input state despite the birefringence.

Alice starts by sending a sequence of photons with a random phase shift $\varphi_{PM1}$ through her interferometer to Bob. The polarization states of the pulses $\Lambda_1$ and $\Lambda_2$ when leaving PBS1 are

$$\left. \begin{array}{l} |\Lambda_1\rangle = \tilde{T}_{SA} \cdot T_{FM\,1} \cdot \vec{T}_{SA} \cdot |v\rangle = -\exp[\,i(\theta_{SA} + \varphi_{PM\,1})] \cdot |h\rangle \\ |\Lambda_2\rangle = \tilde{T}_{LA} \cdot T_{FM\,2} \cdot \vec{T}_{LA} \cdot |h\rangle = -\exp[\,i(\theta_{LA})] \cdot |v\rangle \end{array} \right\} \qquad (2)$$

where $\vec{T}_{SA} = \vec{T}_{PBS1-PM1} \cdot \vec{T}_{PM1} \cdot \vec{T}_{PM1-FM1}$ denotes the forward Jones matrix of the short arm $S_A$ and includes the matrices for the fiber between PBS1 and PM1, PM1, and the fiber between PM1 and FM1, while $\tilde{T}_{SA}$ is the corresponding return matrix; $\vec{T}_{LA} = \vec{T}_{PBS1-FM2}$ is the forward and $\tilde{T}_{LA}$ the backward Jones matrix for the long arm $L_A$; $\theta_{SA}$ and $\theta_{LA}$ are phases due to fiber birefringence in the arms $S_A$ and $L_A$, respectively, and $\varphi_{PM1}$ is the electronically modulated phase shift controlled by Alice. The same corresponding matrices are obtained for Bob's interferometer with the substitution of B for A.

When Bob randomly chooses a modulation phase shift of $\varphi_{PM2}$ the polarization states of the pulses $\Lambda_1$ and $\Lambda_2$ at the output port of his PBS2 become

$$\begin{aligned}|\Lambda'_1\rangle &= \tilde{T}_{LB} \cdot T_{FM4} \cdot \vec{T}_{LB} \cdot \exp[i(\theta_{SA} + \Phi_{PM1})] \cdot (-|h\rangle) = \exp[i(\theta_{SA} + \theta_{LB} + \varphi_{PM1})] \cdot |v\rangle \\ |\Lambda'_2\rangle &= \tilde{T}_{SB} \cdot T_{FM3} \cdot \vec{T}_{SB} \cdot \exp(i\theta_{LA}) \cdot (-|v\rangle) = \exp[i(\theta_{LA} + \theta_{SB} + \varphi_{PM2})] \cdot |h\rangle \end{aligned} \quad (3)$$

The resulting polarization state $\Lambda$ due to superposition of $|\Lambda'_1\rangle$ and $|\Lambda'_2\rangle$ is thus

$$|\Lambda\rangle = \begin{pmatrix} \exp[i(\theta_{LA} + \theta_{SB} + \varphi_{PM2})] \\ \exp[i(\theta_{SA} + \theta_{LB} + \varphi_{PM1})] \end{pmatrix} = \exp(i\theta) \cdot \begin{pmatrix} \exp(i\varphi_{PM2}) \\ \exp(i\varphi_{PM1}) \end{pmatrix} \quad (4)$$

where $\theta = \theta_{LA}+\theta_{SB} = \theta_{SA}+\theta_{LB}$. We thus see that the polarization state of $\Lambda$ is determined by the phases of Alice and Bob's modulators $\varphi_{PM1}$ and $\varphi_{PM2}$. In fact, the original horizontal (vertical) component of Alice's pulse still comes out horizontal (vertical) but carries Bob's (Alice's) phase modulation. For photons injected into Alice's interferometer in the +45° polarization state the final output states corresponding to phase shifts of $\varphi_{PM1}=0$, $\pi/2$, $\pi$ and $3\pi/2$, and $\varphi_{PM2}= 0$, $\pi/2$, $\pi$ and $3\pi/2$ are shown in Table 1, assuming no photons are lost in transmission or detection. By choosing the appropriate form of polarization projection measurement either the BB84 or B92 protocol can be implemented.

If the 4-state BB84 protocol is chosen we use the detection scheme of Fig 1b. For convenience of measurement the output polarization is rotated by a λ/2 wave plate set with its fast axis at an angle of +22.5° with the horizontal axis. A third polarizing beamsplitter PBS3 is used for the projection measurement so that the vertical and horizontal polarization states are detected by D1 and D2, respectively. As shown in Table 1, when Alice and Bob use the same basis (0, π) or (π/2, 3π/2) the photon state can be determined with 100% certainty; with different bases the detection probability is only 50%.

The measurement setup for the B92 protocol is very simple, with just an analyzing polarizer P set at +45° and a detector D, as shown in Fig 1c. Alice can choose $\varphi_{PM1}$= 0 or π/2 for her modulation phase, while Bob can choose $\varphi_{PM2}$= π or 3π/2. Table 1 shows that a photon can pass through the polarizer and be detected with 50% probability when Alice uses a phase of 0 (π/2) and Bob chooses 3π/2 (π). The advantage of this method is that only one single-photon detector is needed.

In order to test the performance of our scheme the experimental setup shown in Figs. 1a and b was used, with a fiber polarization controller as the λ/2 wave plate. Photons generated by a 1.31 μm laser diode (PiLas, Advanced Laser Diode Systems) with a repetition frequency of 1MHz and polarized in the +45° direction were injected after attenuation into Alice's interferometer through PBS1, custom made by Advanced Fiber Resources (Zhuhai), then transmitted through a 50km long quantum channel QC to Bob. The two homemade single-photon detectors D1 and D2 with a dark count of 3 x $10^{-7}$/ns based on EPM239BA InGaAs/InP avalanche photodiodes were synchronized with the laser and triggered at 1 MHz with a gate width of 100ns.

The inset in Fig. 2 shows the counts of D1 and D2 versus the voltage $V_{PM2}$ on Bob's phase modulator which was scanned from 0 to 10V (with PM1 maintained at 0V). The modulators were made by CETC with a half-wave voltage of 5.4V. We note that the count rate of D2 is maximal and D1 minimal when $V_{PM2}$ = 0V, and vice versa when $V_{PM2}$ is 5.4V, corresponding to a relative phase shift $\varphi_{PM2}$-$\varphi_{PM1}$ of 0 or π so that the photons reach PBS3 in the vertical or horizontal polarization state, respectively. Both detectors have the same counts when $V_{PM2}$ is 2.7V and 8.2V, showing that the states are left or right circularly polarized, respectively. The maximum count was 1750/s and the minimum 20/s, the latter being within the dark count rate of the detectors. The polarization extinction ratio, defined as the maximum divided by the minimum counts was about 30dB and could be maintained for several hours, as shown in Fig. 2, without any adjustments to the system. The slight decrease towards the end is due to gradual phase drift in the trunk fiber, which can be corrected for by the polarization controller PC in a practical QKD system.

In summary, we have demonstrated a one-way QKD scheme based on phase encoding and polarization measurement of single photons with intrinsically stable Faraday-mirror type Michelson interferometers which can compensate for birefringence effects automatically. A polarization extinction ratio as high as 30dB has been attained, and the interferometer can remain stable indefinitely, showing its potential for use in practical systems.

This work was supported by the National Program for Basic Research in China Grant No. 2001CB309301, and the National Natural Science Foundation of China Grant No. 60578029.

Figure Captions

Table 1: Polarization states of Λ and detection probabilities for different phase values of Alice and Bob.

Fig.1 Schematic of the phase encoding and polarization measurement QKD system. See text for notation. (a) Phase encoding system with dual 4-port polarizing beamsplitters, with polarization measurement setup for the (b) BB84 and (c) B92 protocol.

Fig. 2 Variation of polarization extinction ratio with time. Inset: single-photon counts v. Bob's modulation voltage $V_{PM2}$ for $V_{PM1} = 0V$.

Table 1

| $\varphi_{PM1}$ | | 0 | | | | $\pi/2$ | | | | $\pi$ | | | | $3\pi/2$ | | |
|---|---|---|---|---|---|---|---|---|---|---|---|---|---|---|---|---|
| $\varphi_{PM2}$ | | 0 | $\pi/2$ | $\pi$ | $3\pi/2$ | 0 | $\pi/2$ | $\pi$ | $3\pi/2$ | 0 | $\pi/2$ | $\pi$ | $3\pi/2$ | 0 | $\pi/2$ | $\pi$ | $3\pi/2$ |
| $\|\Lambda\rangle$ | | ↗ | ↻ | ↙ | ↺ | ↻ | ↗ | ↺ | ↙ | ↙ | ↺ | ↗ | ↻ | ↺ | ↙ | ↻ | ↗ |
| BB84 | D1(%) | 0 | 50 | 100 | 50 | 50 | 0 | 50 | 100 | 100 | 50 | 0 | 50 | 50 | 100 | 50 | 0 |
|  | D2(%) | 100 | 50 | 0 | 50 | 50 | 100 | 50 | 0 | 0 | 50 | 100 | 50 | 50 | 0 | 50 | 100 |
| B92 | D(%) |  |  | 0 | 50 |  |  | 50 | 0 |  |  |  |  |  |  |  |  |

↗ : 45° linear polarization;  ↙ : -45° linear polarization;  ↺ : Left circular polarization;  ↻ : Right circular polarization

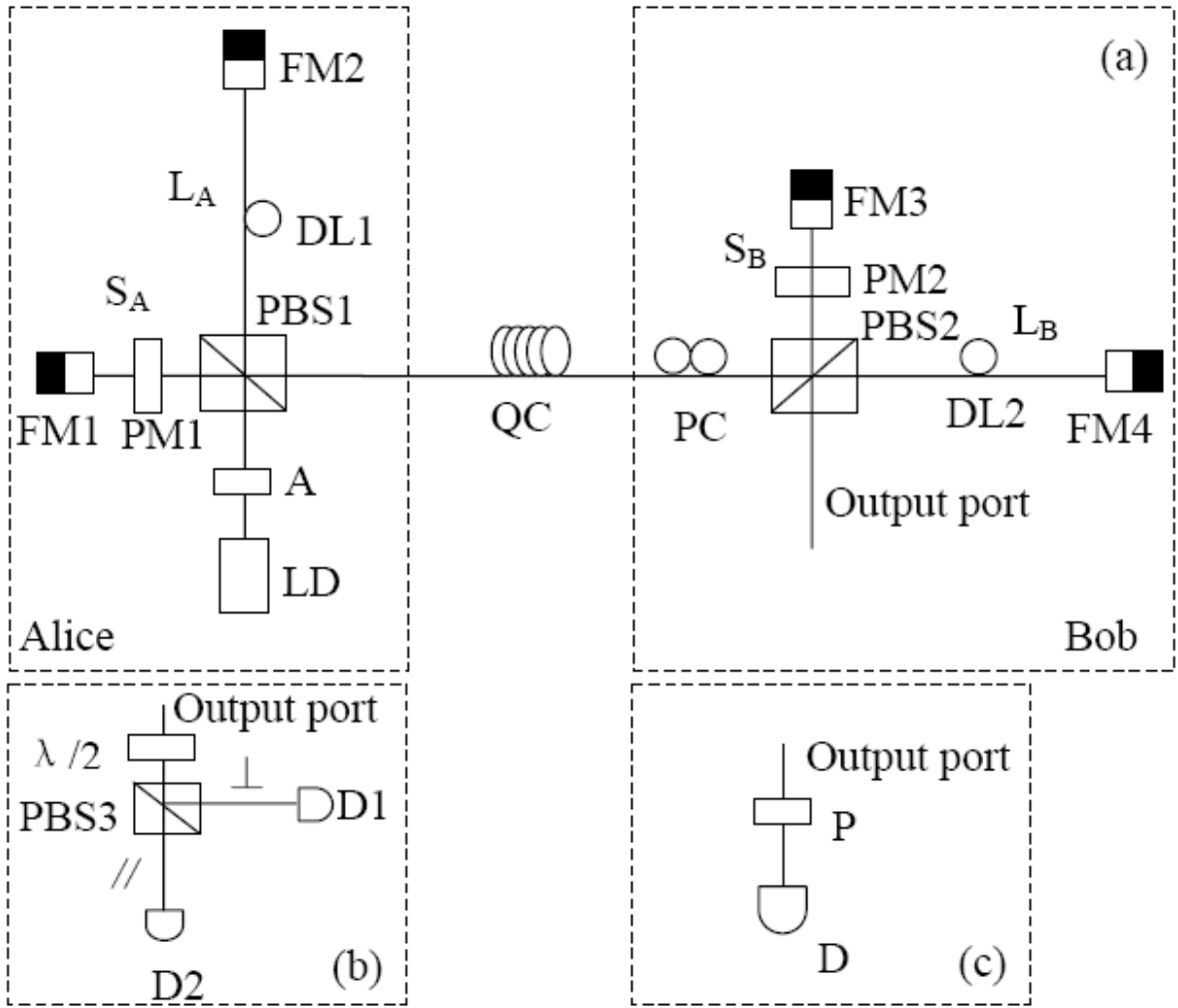

Fig. 1

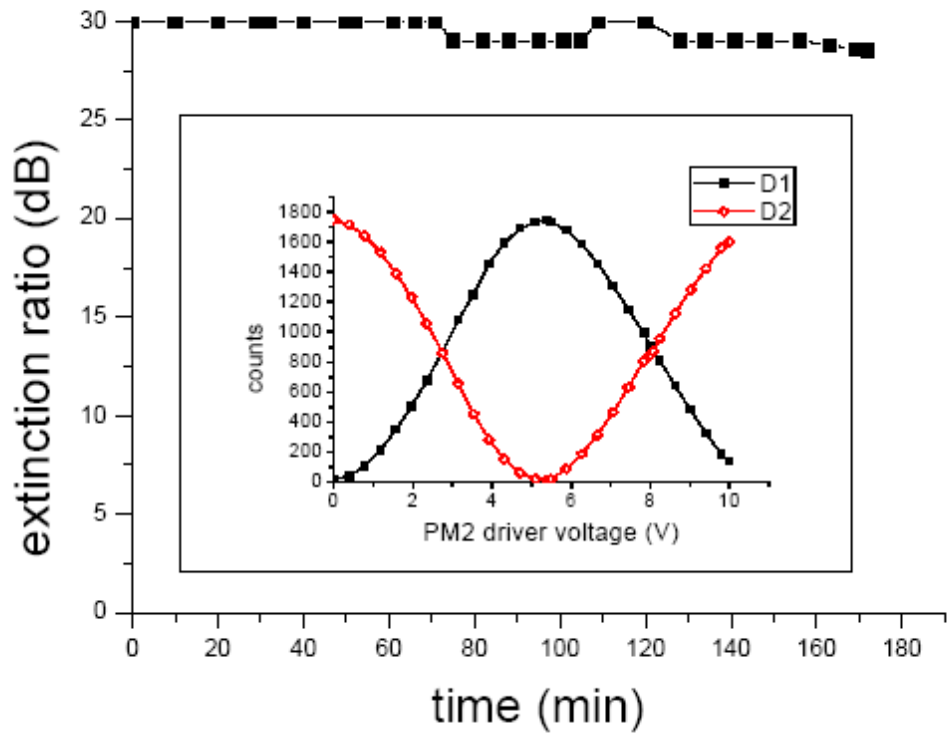

Fig. 2